\documentclass{appolb}

\usepackage{graphicx}
\usepackage{bm} 
\usepackage{subfigure}
\usepackage[T1]{fontenc}

\begin{document}

\title{{\bf Hydrodynamics of transversally thermalized partons in ultra-relativistic heavy-ion collisions}
\thanks{This investigation was partly supported by the Polish Ministry Of Science and Higher Education grants Nos. N202 153 32/4247 and N202 034 32/0918.}}
\author{Mikolaj Chojnacki $^1$  and Wojciech Florkowski $^{1,2}$ 
\address{
$^1$ Institute of Nuclear Physics, Polish Academy of Sciences \\
ul. Radzikowskiego 152, PL-31342 Krak\'ow, Poland \\
$^2$ Institute of Physics, \'Swi\c{e}tokrzyska Academy, \\
ul.~\'Swi\c{e}tokrzyska 15, PL-25406~Kielce,~Poland \\
}}

\maketitle

\begin{abstract}
The hydrodynamic description of transversally thermalized matter, possibly formed at the early stages of ultra-relativistic heavy-ion collisions, is developed. The formalism is based on the thermodynamically consistent approach with all thermodynamic variables referring to two-dimensional objects, the so-called transverse clusters, which are identified with the particles having the same rapidity. The resulting hydrodynamic equations for a single cluster have the form of the two-dimensional hydrodynamic equations of the perfect fluid. Since the clusters do not perform any work in the longitudinal direction, their energy is completely transformed and used to generate strong radial and elliptic flows that turn out to be compatible with the experimental data.
\end{abstract}

\noindent pacs numbers: 25.75.-q, 25.75.Dw, 25.75.Ld

\noindent keywords: ultra-relativistic heavy-ion collisions, relativistic hydrodynamics

\maketitle 

\section{Introduction}
\label{sect:Intro}

The data collected at RHIC indicates that matter produced in ultra-relativistic heavy-ion collisions behaves like an almost perfect fluid \cite{Heinz:2005zg,Shuryak:2005}. This observation triggers many new developments of relativistic hydrodynamics of perfect and viscous fluids \cite{Teaney:2003kp,Hirano:2005wx,Hirano:2005dc,Hirano:2005xf,Hama:2005dz,Eskola:2005ue,Heinz:2005bw,Nonaka:2005aj,Andrade:2006yh,Koide:2006ef,Nonaka:2006yn,Hirano:2007xd,Baier:2006sr,Huovinen:2006jp,Satarov:2006jq,Baier:2006um}. On the other hand, the hydrodynamic picture is still challenged by the two serious problems. The first one refers to an unexpected short thermalization scale that is required to describe the measured asymmetry of the transverse flow, the second one
is connected with the difficulty to explain the measured HBT radii (the so-called HBT puzzle).

Recently, it was proposed that these difficulties can be avoided if at the early stages of the evolution of matter produced in ultra-relativistic heavy-ion collisions the hydrodynamic description applies only to the transverse degrees of freedom while the longitudinal expansion may be to large extent regarded as the motion of independent clusters \cite{my}. In this paper we explain in more detail and develop the concept of Ref. \cite{my}. The idea of purely transverse equilibration has been analyzed previously by Heinz and Wong in Ref. \cite{HeinzWong} with the conclusion that it cannot be realistic since it does not lead to the large elliptic flow found in the corresponding 3-dimensional (3D) hydrodynamic calculations. Our conclusions differ substantially from those reached in \cite{HeinzWong} because of at least two reasons: Firstly, we compare the results of our model calculations to the present data rather than to other hydrodynamic calculations. Secondly, we use a different technical implementation of the concept of transverse thermalization and longitudinal free-streaming.

The formalism introduced in \cite{my} and developed in this paper is based on the thermodynamically consistent approach where all thermodynamic variables refer to two-dimensional objects. We call them transverse clusters and identify with the particles having the same rapidity. The resulting hydrodynamic equations for a single cluster, i.e. for a fixed value of rapidity, have the form of the 2-dimensional (2D) hydrodynamic equations of perfect fluid. Since the clusters do not perform any work in the longitudinal direction, their energy is completely transformed and used to generate strong radial and elliptic flows that turn out to be compatible with the experimental data. In contrast to Ref. \cite{HeinzWong} our approach conserves entropy and, therefore, describes perfect fluid.

An attractive feature of our approach are short space and time scales characterizing the system at the moment when the substantial elliptic flow is formed. We expect that this feature may help to solve the HBT puzzle and the problem of early equilibration. The verification of this point requires, however, further developments, e.g. an implementation of the hadronization scheme which transforms partons into the observed hadrons.

The paper is organized as follows: In Sects. \ref{sect:ansatz} - \ref{sect:HE} we introduce the phase-space distribution function, its moments, and the equations of transverse hydrodynamics, respectively. The moments of the distribution function define the particle density current, the energy-momentum tensor, and the entropy current. The hydrodynamic equations are obtained from the conservation laws for the energy and momentum. In Sects. \ref{sect:initcond} and \ref{sect:freezeout} we discuss our initial conditions and the Cooper-Frye formula. In Sects. \ref{sect:results} and 8 we present our results and conclusions. Finally, in the Appendix the technical details of our calculations are given.

\section{Ansatz for the phase-space distribution function}
\label{sect:ansatz}

\bigskip

\begin{figure}[t]
\begin{center}
\includegraphics[angle=0,width=0.6\textwidth]{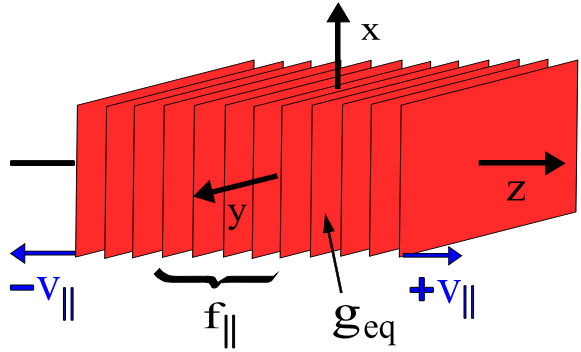}
\end{center}
\caption{Visualization of the ansatz (\ref{fxp}) - (\ref{fparallel}). Particles having the same longitudinal velocity form transverse clusters. Their dynamics is governed by the equations of 2D hydrodynamics of perfect fluid. The rapidity distribution of clusters is defined by the function $f_\parallel$. }
\label{fig:vis}
\end{figure}

In our approach we assume the following factorization of the phase-space distribution function $f(x,p)$ into the longitudinal and transverse part, see Fig. \ref{fig:vis},
\begin{eqnarray}
f(x,p) &=& f_\parallel \, g_{\rm eq} (\tau ,\eta ,%
\overrightarrow{x}_{\bot },\overrightarrow{p}_{\bot })\,,
\label{fxp} 
\end{eqnarray}
where
\begin{eqnarray}
f_\parallel  &=& n_0 \delta(p_{\Vert } t - E z ) =
 n_0 \frac{\delta(\eta - y)}{ \tau m_{\bot} } .
\label{fparallel}
\end{eqnarray}
Here the standard definitions of the energy and longitudinal momentum in terms of the rapidity have been used, 
\begin{equation}
E = m_{\bot} \, \hbox{cosh} y, \quad p_{\Vert } = m_{\bot} \, \hbox{sinh} y.
\label{rapidity}
\end{equation}
Similarly, the spacetime coordinates $t$ and $z$ have been expressed in terms of the proper time $\tau$ and spacetime rapidity $\eta$, 
\begin{equation}
t = \tau \, \hbox{cosh} \eta, \quad z = \tau \, \hbox{sinh} \eta.
\label{spacetimerap}
\end{equation}
We note that the form of the distribution (\ref{fxp}) - (\ref{fparallel})  is very much similar to that used by Heinz and Wong in Ref. \cite{HeinzWong}, the only formal difference is that the factor $n_0/m_{\bot}$ in Eq. (\ref{fxp}) is replaced in Ref. \cite{HeinzWong} by the factor $\tau_0$. With our ansatz all thermodynamic variables which appear in the formalism are consistently defined as two-dimensional quantities and the resulting equations describe perfect fluid. This is different from Ref. \cite{HeinzWong} where the authors treat their approach as non-equilibrium viscous dynamics which generates entropy.

In Eq. (\ref{fparallel}) the dimensionless normalization parameter $n_0$ describes the density of clusters in rapidity. The role of this parameter may be compared with the role played by the time scale parameter $\tau_0$ introduced in Ref. \cite{HeinzWong}. Heinz and Wong interpret $\tau_0$ as the initial time for the hydrodynamic evolution. A closer inspection of their approach indicates, however, that this parameter plays a role of the global normalization having no relation with the initial time. The hydrodynamic equations of Ref. \cite{HeinzWong} do not contain the $\tau_0$ parameter. The hydrodynamic evolution may start at arbitrary value of the initial time, let us call it $\tau=\tau_{\rm init}$, and the final results are independent of the choice that has been made for $\tau_{\rm init}$. 

The equilibrium distribution $ g_{\rm eq} $ has the form of the two-dimensional equilibrium distribution function convoluted with the transverse flow. For simplicity we use the Boltzmann statistics and neglect the chemical potential. In this case we have 
\begin{equation}
g_{\rm eq} = \exp \left( -\frac{m_{\bot }\,u_{0}-\overrightarrow{p}_{\bot }\cdot 
\overrightarrow{u}_{\bot }}{T}\right),
\label{geq}
\end{equation}
where the transverse mass $m_{\bot}$ is defined by the formula $m_{\bot} = \sqrt{m^2+p^2_{\bot}}$. The transverse flow $u^\mu$ has the structure
\begin{eqnarray}
& & u^{\mu } =
\left( u_{0},\,u_{x},\,u_{y},0\right) =\left( u_{0},\,\overrightarrow{u}_{\bot },0\right),
\nonumber \\ 
& & u_{0}^{2}-\overrightarrow{u}_{\bot }^{2}=1, 
\label{smallu}
\end{eqnarray}
which explicitly underlines its transverse character (vanishing longitudinal component, $u_z=0$).

\section{Moments of the phase-space distribution function}
\label{sect:moments}

By calculating the appropriate momentum integrals of the distribution function one obtains the particle current $N^\mu$, the energy-momentum tensor $T^{\mu \nu}$, and the entropy current $S^\mu$. Treating particles as massless with the ansatz (\ref{fxp}) one finds
\begin{eqnarray}
N^\mu &=& n_0 \nu_g \int \frac{dy\, d^{2}p_{\bot \,\,}}{(2\pi )^{2}} p^\mu 
\,\frac{\delta(y-\eta)}{\tau m_{\bot }}g_{\rm eq} 
\nonumber \\ 
&=& \frac{n_0 \nu_g  T^2}{2\pi \tau} U^\mu, \label{Nmu}  \\
T^{\mu \nu} &=& n_0 \nu_g  \int \frac{dy\, d^{2}p_{\bot \,\,}}{(2\pi )^{2}} p^\mu p^\nu 
\,\frac{\delta(y-\eta)}{\tau m_{\bot }}g_{\rm eq}\nonumber \\
&=& \frac{n_0 \nu_g  T^3}{2\pi \tau} \left( 3 U^\mu U^\nu -g^{\mu \nu} - V^\mu V^\nu \right), 
\label{Tmunu}  \\
S^\mu &=&  -n_0 \nu_g \int  \frac{dy\, d^{2}p_{\bot \,\,}}{(2\pi )^{2}}p^{\mu }
\,\frac{\delta(y-\eta)}{\tau m_{\bot }}g_{\rm eq} \left( \ln g_{\rm eq} - 1\right) 
\nonumber \\
&=& \frac{3 n_0 \nu_g  T^2}{2\pi \tau} U^\mu.   \label{Smu}
\end{eqnarray} 
Here we have introduced the degeneracy factor $\nu_g$ connected with the density of states in the transverse space. Treating our system as dominated by gluons we assume that $\nu_g = 16$. In Eqs. \mbox{(\ref{Nmu}) - (\ref{Smu})} the 3D hydrodynamic flow $U^\mu$  has the  structure
\begin{equation}
U^{\mu } =\left( \cosh \eta \,u_{0},\,\,\,\,u_{x},\,\,u_{y},\,\sinh \eta
\,u_{0}\right).
\label{bigu}
\end{equation}
We note that the correct normalization of the transverse flow $u^\mu$ leads directly to the correct normalization of the four-vector $U^\mu$, namely $U^\mu U_\mu = 1$. We also note that the energy-momentum tensor includes an extra contribution which does not appear in the standard hydrodynamics. The extra term is proportional to the product $V^\mu V^\nu$, where the four vector $V^\mu$ has the structure
\begin{equation}
V^{\mu } =\left( \sinh \eta,\,\,\,\,0,\,\,0,\,\cosh \eta \right).
\label{bigv}
\end{equation}
The four vector $V^\mu$ is spacelike, $V^\mu V_\mu = -1$, and its origin is connected with special role of the longitudinal direction in our case, at $\eta=0$ it takes the form $V^\mu = (0,0,0,1)$. With the presence of the term proportional to $V^\mu V^\nu$, the energy-momentum tensor is traceless, \mbox{$T^\mu_{\,\,\,\mu} = 0$}, as required for massless particles. The matrix form of the energy-momentum tensor is given in the Appendix.

The appearance of the extra terms in the energy-momentum tensor was already discussed in Ref. \cite{HeinzWong}. We stress that our structure is simpler than that discussed by Heinz and Wong and this fact is connected with the identification of the two-dimensional thermodynamic properties of the system. With our ansatz for the distribution function, the time components of the currents  (\ref{Nmu}) - (\ref{Smu}) reduce in the rest frame of the system to the two-dimensional thermodynamic densities divided by the proper time. In particular,  in the frame where $U^\mu = (1,0,0,0)$ we find
\begin{eqnarray}
N^0 =\frac{n_0}{\tau} \, n_2, \quad T^{00} = \frac{n_0}{\tau} \, \varepsilon_2, 
\quad S^0 = \frac{n_0}{\tau} \,s_2,
\label{zerocomp}
\end{eqnarray}
where the appropriate two-dimensional densities are defined by the equations
\begin{eqnarray}
n_{2} &=& \nu_g \int \frac{d^{2}p}{\left( 2\pi \right) ^{2}} g_{\rm eq} =  \frac{\nu_g T^2}{2\pi}, 
\nonumber \\
\varepsilon _{2} &=& \nu_g  \int \frac{d^{2}p}{\left( 2\pi \right) ^{2}}\, p_\bot g_{\rm eq} =  \frac{\nu_g T^3}{\pi},  \nonumber \\
s_{2} &=& -\nu_g  \int \frac{d^{2}p}{%
\left( 2\pi \right) ^{2}}g_{eq}\left( \,\ln g_{eq}-1\right) =  \frac{3 \nu_g T^2}{2\pi}. 
\label{2dens}
\end{eqnarray}
Those definitions may be supplemented by the definition of pressure, which is obtained from the thermodynamic relation
\begin{equation}
\varepsilon_2 + P_2 = T s_2.
\label{epsP}
\end{equation}
A simple calculation gives
\begin{equation}
P_2 =  \frac{\nu_g T^3}{2\pi} =  n_2 T = \frac{\varepsilon_2}{2}.
\label{pressure}
\end{equation}
The last equality, in agreement with the expectations for two-dimensional systems, yields the sound velocity
\begin{equation}
c_s^2 = \frac{1}{2}.
\label{cs2}
\end{equation}

We note that the factor $1/\tau$ in Eqs. (\ref{zerocomp}) is of pure kinematic origin and describes the decrease of three dimensional densities due to the increasing distance between the transverse clusters. At midrapidity we have $dz = \tau \,dy$ and the rapidity densities per unit transverse area,  $dA=dxdy$, are
\begin{eqnarray}
\frac{dN}{dA dy} = n_0 n_2, \quad 
\frac{dE}{dA dy} = n_0 \varepsilon_2, \quad 
\frac{dS}{dA dy} = n_0 s_2.
\label{zerocomp1}
\end{eqnarray}
Eqs. (\ref{zerocomp1}) are natural for 2D systems, and one can see that rapidity densities per unit transverse area  change only if the temperature decreases, i.e., only if the transverse flow is present. We note that this type of behavior is completely different from the scenario assumed in the Bjorken model.

The temperature dependence of our 2D thermodynamic variables is different from the scaling of the energy density $\varepsilon_3 \sim T^4$ and pressure $P_3 \sim T^4$ that was used in Ref. \cite{HeinzWong}. The relations $\varepsilon_3 \sim T^4, P_3 \sim T^4$, and $\varepsilon_3 = 2 P_3$, {\it when used in equilibrium}, lead to the contradictory results for the value of the sound velocity. The first two yield $s_3 \sim T^3$ and $c_s^2 = 1/3$, whereas the third one gives $c_s^2 = 1/2$.  Consequently, the approach presented in Ref. \cite{HeinzWong} should be interpreted as an effective description of viscous dynamics, while in our approach the transverse dynamics is reduced to 2D hydrodynamics of perfect fluid with $c_s^2 = 1/2$.

The large value of the sound velocity leads, as expected, to much stiffer than usual equation of state, and favors formation of the strong elliptic flow. The most important effect responsible for the formation of the strong flow is, however, the lack of the interaction between the transverse clusters. Since the energy of the clusters is not reduced by the work done in the longitudinal direction, it is exclusively transformed and used to generate strong radial and elliptic flows.

\section{Transverse hydrodynamics}
\label{sect:HE}

The hydrodynamic equations are obtained from the energy and momentum conservation laws
\begin{equation}
\partial_\mu T^{\mu \nu} = 0,
\label{enmomcons}
\end{equation}
where the energy-momentum tensor is given by Eq. (\ref{Tmunu}). We have checked that Eqs. (\ref{enmomcons}) imply  the entropy conservation law
\begin{equation}
\partial_\mu S^\mu = 0,
\label{entrcons}
\end{equation}
with the entropy current defined by Eq. (\ref{Smu}). The connection between Eqs. (\ref{enmomcons}) and Eq. (\ref{entrcons}) is the same as in the standard 3D hydrodynamics of perfect fluid where the entropy conservation law $\partial_\mu S^\mu = 0$ is obtained by the projection of the energy-momentum conservation laws on the four-velocity of the fluid, i.e., from the formula $U_\nu \partial_\mu T^{\mu \nu} = 0$. 

\begin{figure}[t]
\begin{center}
\includegraphics[angle=0,width=0.45\textwidth]{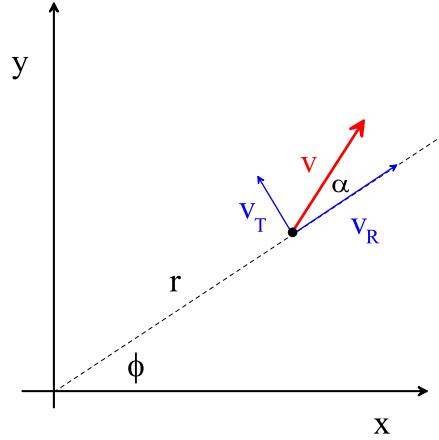}
\end{center}
\caption{Geometrical definition of the dynamical angle $\alpha$. In our approach, instead of the two components of the transverse flow we use the magnitude of the flow $v$ and the angle $\alpha$. Both $v$ and $\alpha$ are functions of $\tau, r$ and $\phi$. }
\label{fig:alpha}
\end{figure}

The use of our form of the energy-momentum tensor in Eq. (\ref{enmomcons}) leads to the following three equations
\begin{eqnarray}
\frac{\partial }{\partial \tau} \left( r s_2 u_0 \right) +\frac{\partial }{\partial r}
\left( r s_2 u_0 v\cos \alpha \right) + \frac{\partial }{\partial
\phi }\left( s_2 u_0 v\sin \alpha \right)  &=&0, 
\nonumber \\
\frac{\partial }{\partial \tau}\left( rTu_0 v\right) +
r\cos \alpha \frac{\partial }{\partial r}\left( Tu_0 \right) 
+\sin \alpha \frac{\partial }{\partial \phi }\left( Tu_0 \right)  &=&0, 
\nonumber \\
Tu_0 ^{2}v\left( \frac{d\alpha }{d \tau}+\frac{v\sin \alpha }{r}\right) -\sin
\alpha \frac{\partial T}{\partial r}
+\frac{\cos \alpha }{r}\frac{\partial T}{\partial \phi } &=&0. \nonumber \\
\label{wfdyr}
\end{eqnarray}
The first equation in (\ref{wfdyr}) is just the entropy conservation, compare Eq. ({\ref{entrcons}). The second equation in (\ref{wfdyr}) follows from the symmetric linear combination \mbox{$U_1 \partial_\mu T^{\mu 1}+U_2 \partial_\mu T^{\mu 2} = 0$}, while the third equation results from the asymmetric linear combination  \mbox{$U_2 \partial_\mu T^{\mu 1} - U_1 \partial_\mu T^{\mu 2} = 0$}. In Eqs. (\ref{wfdyr}) we used the cylindrical coordinates 
\begin{equation}
r\!\!=\!\!\sqrt{x^2+y^2}, \quad \phi=\hbox{tan}^{-1} (y/x).
\label{cylinder}
\end{equation}
The quantity $v$ is the transverse flow, $u_0 =\left(1-v^2\right)^{-\frac{1}{2}}$,  and $\alpha$ is the dynamical angle describing deviation of the direction of the flow from the radial direction, 
\begin{equation}
v_x = v \cos(\alpha+\phi), \quad v_y = v \sin(\alpha+\phi),
\label{vxvy}
\end{equation}
see Fig. \ref{fig:alpha}. The deferential operator ${d}/{d\tau}$ is the total time derivative defined by the formula
\begin{equation}
\frac{d}{d\tau} = \frac{\partial}{\partial \tau} 
+  v \cos\alpha \frac{\partial}{\partial r} 
+ \frac{v \sin\alpha}{r} \frac{\partial}{\partial \phi}.
\label{fullddt}
\end{equation} 
The structure of Eqs. (\ref{wfdyr}) is  similar to the structure of the equations describing the boost-invariant 3D hydrodynamic expansion, see \cite{Dyrek:1984xz} and \cite{Chojnacki:2006tv}, 
\begin{eqnarray}
\frac{\partial }{\partial \tau} \left(r s_3 u_0 \right) \!+\!
\frac{\partial }{\partial r}\left( r  s_3 u_0 v\cos \alpha \right) \!+\! 
\frac{\partial }{\partial \phi }\left(s_3 u_0 v\sin \alpha \right) \! &=& \! - \frac{r s_3 u_0 }{\tau} , 
\nonumber \\
\frac{\partial }{\partial \tau}\left( rTu_0 v\right) +
r\cos \alpha \frac{\partial }{\partial r}\left( Tu_0 \right) 
+\sin \alpha \frac{\partial }{\partial \phi }\left( Tu_0 \right)  &=&0, 
\nonumber \\
Tu_0 ^{2}v\left( \frac{d\alpha }{d \tau}+\frac{v\sin \alpha }{r}\right) -\sin
\alpha \frac{\partial T}{\partial r}
+\frac{\cos \alpha }{r}\frac{\partial T}{\partial \phi } &=&0. \nonumber \\
\label{3Dbinv}
\end{eqnarray}
They differ from Eqs. (\ref{wfdyr}) by the presence of the term $r s_3 u_0/\tau$ on the right-hand-side of the first equation in (\ref{3Dbinv}) and by the different form of the entropy density. We recall that the three dimensional entropy density $s_3$ of massless particles with $\nu$ internal degrees of freedom is given by the expression 
\begin{equation}
s_3 = \frac{2}{45} \nu \pi^2 T^3. 
\label{s3}
\end{equation} 
The main physical difference between Eqs. (\ref{wfdyr}) and (\ref{3Dbinv}) resides in the term $(r s_3 u_0)/\tau$ leading to the decrease of the entropy and energy densities even in the case where the transverse flow is absent. The physical origin of this term is the presence of the work which is done in the longitudinal direction. In the case of the 2D expansion such a term is not present and the deposited collision energy is used only to produce the transverse flow.   

We note that (\ref{wfdyr}) is valid even if the temperature $T$, the transverse flow $v$, its direction $\alpha$, and the normalization factor $n_0$ depend on the spacetime rapidity $\eta$. The  structure of the energy-momentum tensor (\ref{Tmunu}) implies   that all partial derivatives with respect to $\eta$ are multiplied by the expressions which vanish in the case $\eta = y$. Consequently, Eqs. (\ref{wfdyr}) are valid for any value of the rapidity, and they should be solved, with the corresponding initial condition, independently for each value of $\eta$. This property shows explicitly that our system is not boost-invariant and may indeed be treated as a superposition of the independent transverse clusters.

The change of the initial entropy density is nicely described by the global conservation laws which follow from Eqs. (\ref{wfdyr}) and (\ref{3Dbinv}). The integration of the first equation in (\ref{wfdyr}) over the transverse spacetime coordinates yields
\begin{equation}
\int\limits_0^\infty dr \, r \int\limits_0^{2 \pi} d\phi \, T^2(\tau,r,\phi) u_0(\tau,r,\phi) 
= \hbox{const}.
\label{glob1}
\end{equation}
On the other hand, for the 3D boost-invariant case one finds
\begin{equation}
\int\limits_0^\infty dr \, r \int\limits_0^{2 \pi} d\phi \, T^3(\tau,r,\phi) u_0(\tau,r,\phi) 
= \frac{\hbox{const}}{\tau}.
\label{glob2}
\end{equation}
For 2D hydrodynamic expansion, see Eq. (\ref{glob1}), the initial thermal energy may decrease only at the expense of increasing transverse flow. For 3D boost-invariant expansion, see Eq. (\ref{glob2}), even without the transverse expansion the temperature drops, as is well known from the Bjorken model. 

We solve Eqs. (\ref{wfdyr}), and also for the comparison Eqs. (\ref{3Dbinv}), using the technique presented in Ref. \cite{Chojnacki:2006tv} which is a direct extension of the method proposed in Ref. \cite{Baym:1983sr}. This method satisfies very accurately the conservation laws (\ref{glob1}) and (\ref{glob2}). We find this agreement as an important check of our numerical scheme.

\section{Initial conditions}
\label{sect:initcond}

The hydrodynamic equations (\ref{wfdyr}) are three equations for three unknown functions: $T$, $v$, and $\alpha$. At the initial time $\tau=\tau_{\rm init}$ the transverse flow $v$ is zero. We also set the angle $\alpha$ to be zero at $\tau=\tau_{\rm init}$. On the other hand the temperature profile at $\tau=\tau_{\rm init}$ is not trivial and its asymmetry in the transverse plane generates the elliptic flow. 

Similarly to other hydrodynamic calculations we assume that the initial energy density
at the transverse position point $\overrightarrow{x}_{\bot } $ is proportional to the wounded-nucleon density $\rho_{WN}$ at this point, namely 
\begin{equation}
\varepsilon_2 \left( \overrightarrow{x}_{\bot } \right) =
\frac{\nu T^3\left( \overrightarrow{x}_{\bot } \right) }{\pi}  \propto 
 \rho_{WN} \left( \overrightarrow{x}_{\bot } \right) . 
\label{initialeps2}
\end{equation}
We note that the assumption (\ref{initialeps2}) used by us for a 2D system is equivalent to the assumption $s_3 \propto   \rho_{WN}$ used in 3D hydrodynamic codes. In our practical calculations Eq. (\ref{initialeps2}) takes the form
\begin{equation}
T(\tau_{\rm init},\overrightarrow{x}_{\bot } ) = T_i \left[
\frac{ \rho_{WN} \left( \overrightarrow{x}_{\bot } \right) }{\rho_{WN} \left(0 \right)}
\right]^{1/3},
\label{Tt0}
\end{equation}
where the parameter $T_i$ is the initial central temperature and the wounded-nucleon density is obtained from the formula \cite{bbc}
\begin{eqnarray}
& &\!\!\!\!\!\!\!\! \rho_{WN} \left( \overrightarrow{x}_{\bot } \right) = \nonumber \\
& & 
T_A\left(\frac{ \overrightarrow{b} }{2}+ { \overrightarrow{x}_{\bot } } \right)
\left\{1\! -\! \left[1\! -\! \frac{\sigma}{A} \, 
T_A\left(-\frac{\overrightarrow{b} }{2}+ {\overrightarrow{x}_{\bot } } \right)
\right]^A \right\} \nonumber \\
& &\!\!\!\!\!\!\!\! + \, T_A\left(-\frac{\overrightarrow{b} }{2}+ {\overrightarrow{x}_{\bot } } \right)
\left\{1\! -\! \left[1\! -\! \frac{\sigma}{A} \, 
T_A\left(\frac{\overrightarrow{b} }{2}+ {\overrightarrow{x}_{\bot } } \right)
\right]^A \right\}. \nonumber \\
\label{dNp}
\end{eqnarray}
In Eq. (\ref{dNp}) $\sigma$ = 40 mb is the total nucleon-nucleon cross section and $T_A\left(x,y\right)$ is the nucleus thickness function 
\begin{equation}
T_A(x,y) = \int dz \, \rho\left(x,y,z\right).
\label{TA}
\end{equation}
Here $\rho(r)$ is the nuclear density profile given by the Woods-Saxon function with the conventional choice of  parameters used for the gold nucleus: 
\begin{eqnarray}
\rho_0 &=& 0.17 \,\hbox{fm} ^{-3},  \nonumber \\
r_0 &=& (1.12 A^{1/3} -0.86 A^{-1/3}) \,\hbox{fm},  \nonumber \\
a &=& 0.54 \,\hbox{fm}, \quad A = 197.
\label{woodssaxon}
\end{eqnarray}
The value of the impact parameter in Eq. (\ref{dNp}) depends on the centrality class considered in the calculations.

\begin{figure}[t]
\begin{center}
\includegraphics[angle=0,width=0.35\textwidth]{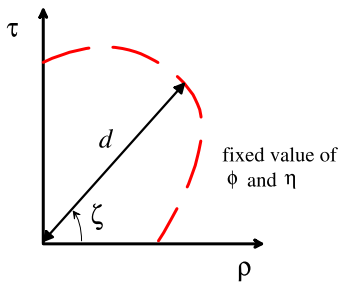}
\end{center}
\caption{Geometric interpretation of the parameters used to define the shape of the hypersurface corresponding to a constant value of the final temperature, $T=T_f$.}
\label{fig:zeta}
\end{figure}

\medskip

\section{Cooper-Frye formula}
\label{sect:freezeout}

To calculate the transverse-momentum spectra at a certain value of the final temperature $T_f$ \footnote{Since we concentrate the evolution of partons this temperature cannot be directly interpreted as the freeze-out temperature.} we use the standard Cooper-Frye prescription
\begin{equation}
\frac{dN}{dy d^2 p_\bot} = \frac{n_0 \nu_g}{(2\pi)^2}\int d\Sigma^\mu p_\mu 
\frac{\delta(\eta - y)}{ \tau m_{\bot} }\, g_{\rm eq},
\label{cooperfrye}
\end{equation}
where the hypersurface $\Sigma$ is defined by the condition $T_f$ = const. For cylindrically asymmetric collisions and midrapidity, $y=0$, the transverse-momentum spectrum has the following expansion in the azimuthal angle of the emitted particles
\begin{equation}
\frac{dN}{dy d^2 p_\bot} = \frac{dN}{dy \, 2 \pi p_\bot \, d p_\bot }
\left( 1 + 2 v_2(p_\bot)  \cos(2 \phi_p) + ...\right).
\label{v2def}
\end{equation}
Eq. (\ref{v2def}) defines the elliptic flow coefficient $v_2$, which may be calculated from (\ref{v2def}) as the asymmetry of the momentum spectrum, 
\begin{eqnarray}
v_2(p_\bot)  &=& \frac{1}{2} 
\frac{ f_N(p_\bot,\phi_p=0) - f_N(p_\bot,\phi_p=\frac{\pi}{2}) }
{ f_N(p_\bot,\phi_p=0) + f_N(p_\bot,\phi_p=\frac{\pi}{2}) },
\nonumber \\
\label{v2cal}
\end{eqnarray}
with $f_N$ being a shorthand notation for $dN/(dy d^2 p_\bot)$.

We are of the opinion that for the essentially 2D expansion, the freeze-out criterion cannot involve the 3D energy density as it was proposed in \cite{HeinzWong}. Application of a 3D freeze-out criterion to a superposition of 2D systems implies that the freeze-out is triggered, in the artificial way, by the increase of the relative distance between the 2D parts. With those remarks in mind we adopt a different strategy; we present our results for different final temperatures and check if the experimental data can be successfully reproduced for one of them.

In our calculations we use the following parameterization of the hypersurface $\Sigma$,
\begin{eqnarray}
t &=& d \left( \phi ,\zeta, \eta \right) \sin \zeta \,\cosh \eta ,\quad
z= d\left( \phi ,\zeta, \eta \right) \sin \zeta \, \sinh \eta , \nonumber \\
x &=& d\left( \phi ,\zeta, \eta\right) \cos \zeta \, \cos \phi ,\quad
y = d\left( \phi ,\zeta, \eta \right) \cos \zeta \, \sin \phi, \nonumber \\
\label{cooperfryeparam}
\end{eqnarray}
which also yields
\begin{eqnarray}
\tau &=& \sqrt{t^2 - z^2} = \,d\left( \phi ,\zeta, \eta \right) \sin \zeta , \nonumber \\
\rho &=& \sqrt{x^2+y^2}= \,d\left( \phi ,\zeta, \eta \right) \cos \zeta  .
\label{tauandrho}
\end{eqnarray}
At any given value of the spacetime rapidity $\eta$ the position of the point on the hypersurface is defined by the two angles, $\phi $ and $\zeta$, and the distance to the origin of the coordinate system, $d\left(\phi ,\zeta, \eta \right)$, see Fig. \ref{fig:zeta}. The angle $\phi $ is the standard azimuthal angle in the $y-x$ plane, while the angle $\zeta$ is the azimuthal angle in the $\tau-\rho$ plane. With the standard definition of the four-momentum in terms of the rapidity and transverse momentum, and with the standard definition of the element of the hypersurface $d\Sigma _{\mu }$ in terms of the totally antisymmetric tensor $\varepsilon _{\mu \alpha \beta \gamma }$, namely

\begin{figure}[t]
\begin{center}
\subfigure{\includegraphics[angle=0,width=0.95\textwidth]{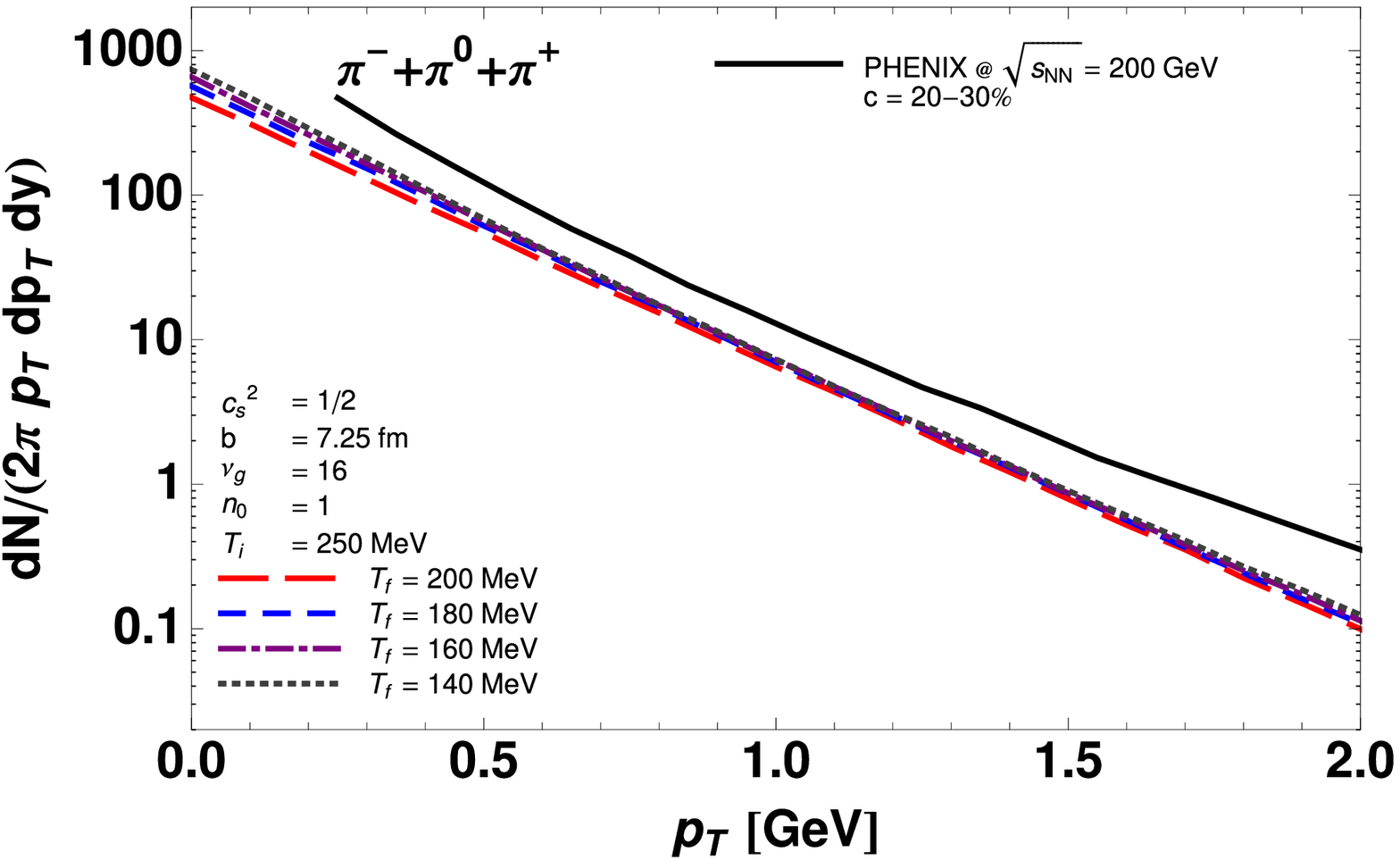}} \\
\subfigure{\includegraphics[angle=0,width=0.95\textwidth]{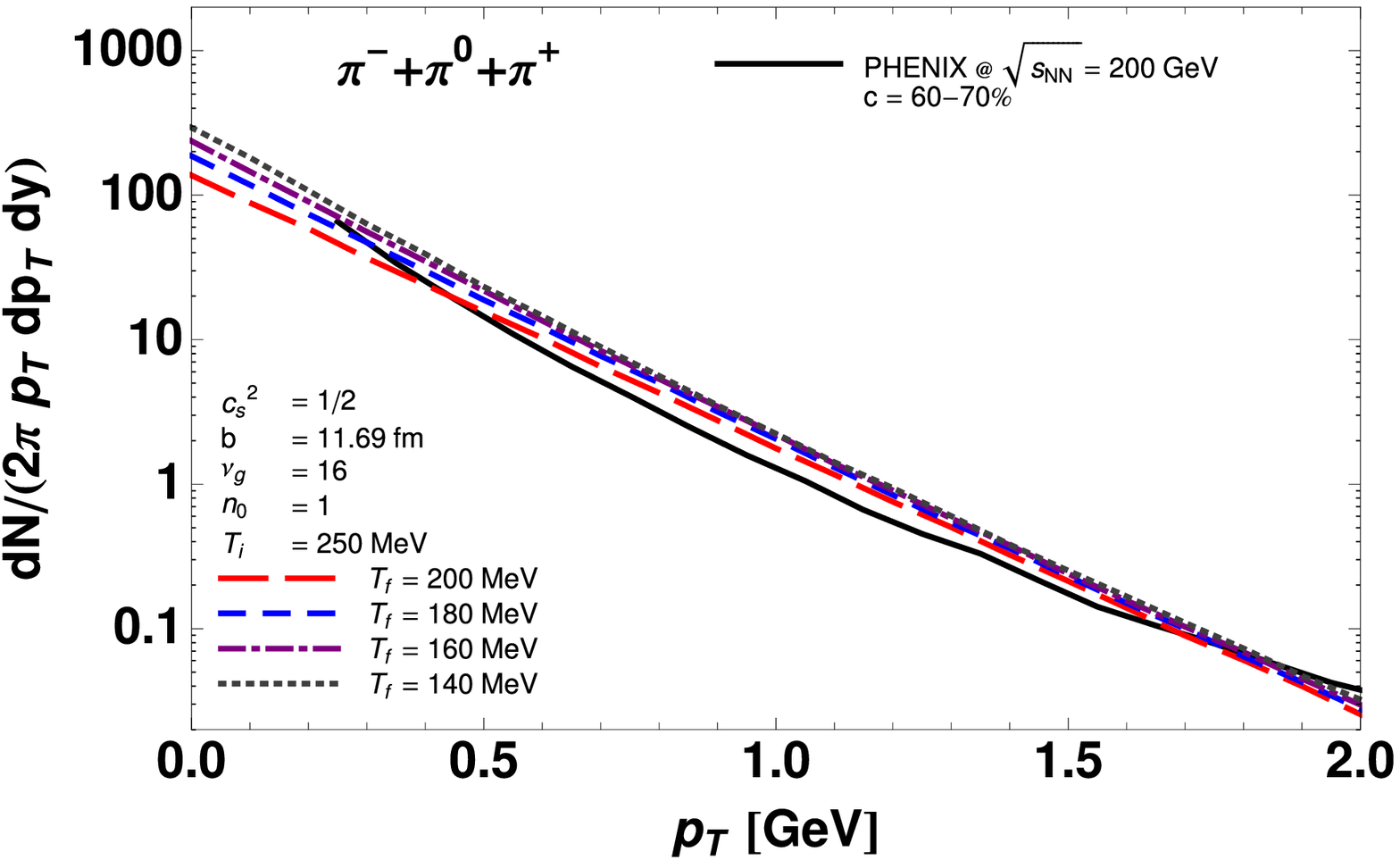}}
\end{center}
\caption{Transverse-momentum spectra for the centrality \mbox{$c$ = 20-30\%} (upper part) and \mbox{$c$ = 60-70\%} (lower part). The PHENIX experimental results \cite{RHIC-spectra} (solid lines) are compared to the model calculations (dashed lines). The initial central temperature, $T_i$ = 250 MeV, is the same in all cases. The four different values of the final temperature are considered: $T_f$ = 200, 180, 160 and 140 MeV. As discussed in the text in more detail, the spectra are to large extent independent of the final temperature (the dashed lines overlap). The normalization factor \mbox{$n_0$ = 1} was used.   }
\label{fig:pT}
\end{figure}

\begin{figure}[t]
\begin{center}
\subfigure{\includegraphics[angle=0,width=0.9\textwidth]{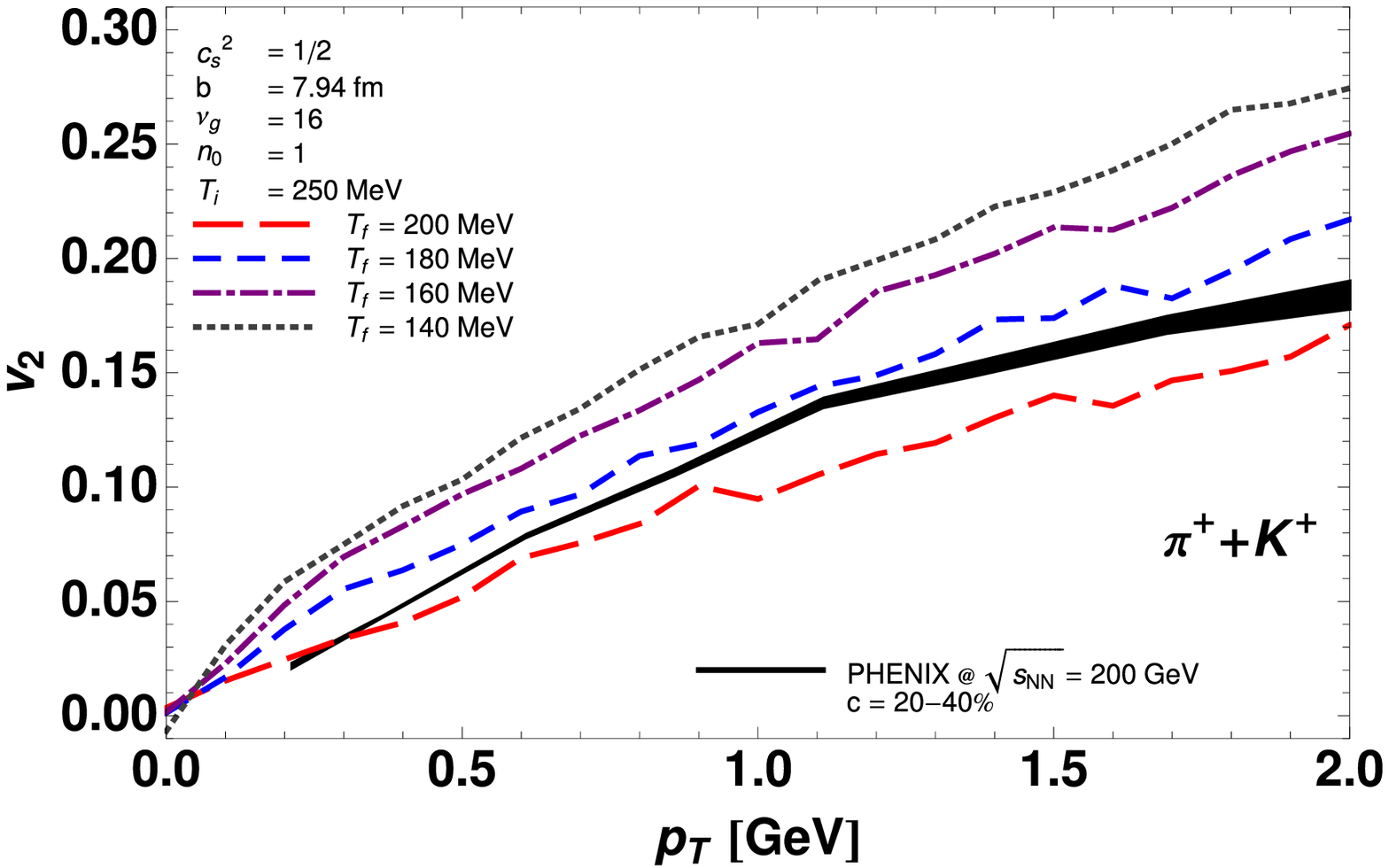}} \\
\subfigure{\includegraphics[angle=0,width=0.9\textwidth]{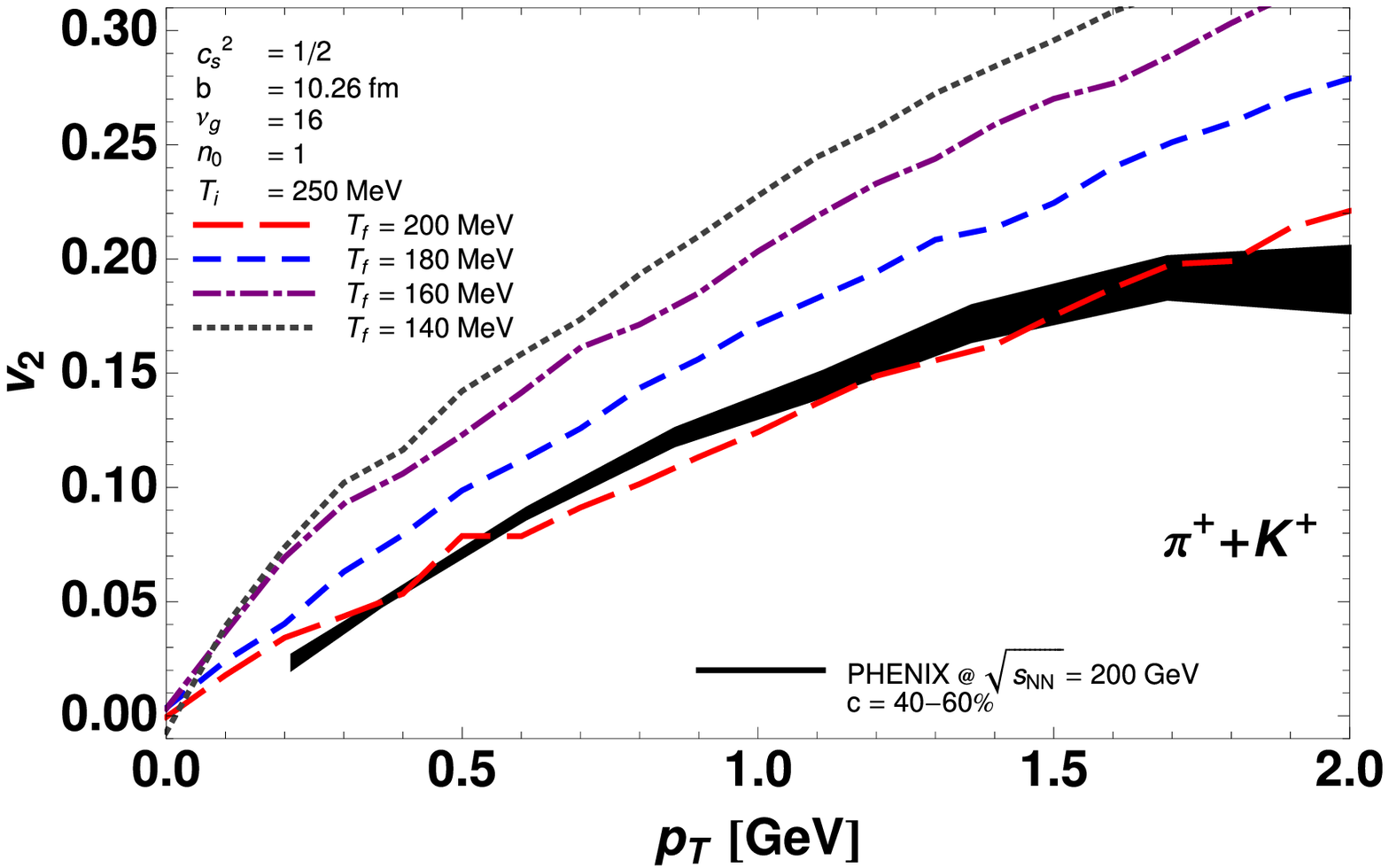}}
\end{center}
\caption{The elliptic flow coefficient $v_2$ for the centrality \mbox{$c$ = 20-40\%} (upper part) and \mbox{$c$ = 40-60\%} (lower part). The solid lines represent the PHENIX data for pions and kaons \cite{RHIC-v2} (the width indicates the experimental error). The dashed lines show our model results. The values of the initial and final temperatures are the same as in Fig. 4.  }
\label{fig:v2}
\end{figure}

\begin{equation}
d\Sigma _{\mu } = \varepsilon _{\mu \alpha \beta \gamma }
\frac{dx^{\alpha }}{d\eta }
\frac{dx^{\beta }}{d\phi }
\frac{dx^{\gamma }}{d\zeta } \,d\eta \,d\phi\, d\zeta,
\label{dSigma}
\end{equation}
we find the explicit form of the Cooper-Frye integration measure
\begin{eqnarray}
d\Sigma _{\mu }\,p^{\mu } & = & d^{\,2}\sin \zeta \left[ 
\vphantom{\frac{\partial d}{\partial \eta } }
d\cos \zeta
\,\left( m_{\bot }\sin \zeta \cosh \left( \eta -y\right) +p_{\bot }\cos
\zeta \cos \left( \phi -\phi _{p}\right) \right) \right.  \nonumber \\
& & +\frac{\partial d}{\partial \zeta }\cos \zeta \,\left( -m_{\bot
}\cos \zeta \cosh \left( \eta -y\right) +p_{\bot }\sin \zeta \cos
\left( \phi -\phi _{p}\right) \right)  \nonumber \\
& & + \frac{\partial d}{\partial \phi }  p_{\bot }\sin \left( \phi -\phi
_{p}\right) \nonumber \\
& &  \left. + \frac{\partial d}{\partial \eta } \, \cot \zeta \,
 m_\bot \sinh \left( \eta - y\right)
\right] d\eta d\phi d\zeta .
\label{dSigma1}
\end{eqnarray}
Eq. (\ref{tauandrho}) and the presence of the delta function in Eq. (\ref{cooperfrye}) imply that the integration measure for massless particles appears always in a simple combination

\begin{eqnarray}
\frac{d\Sigma _{\mu }\,p^{\mu }}{\tau p_{\bot }}\delta \left( \eta -y\right)
&=&d\left[ \vphantom{\frac{\partial d}{\partial \eta } }
d\cos \zeta \,\left( \sin \zeta +\cos \zeta \cos
\left( \phi -\phi _{p}\right) \right) \right.  \nonumber \\
&&+\frac{\partial d}{\partial \zeta }\cos \zeta \,\left( -\cos
\zeta +\sin \zeta \cos \left( \phi -\phi _{p}\right) \right)  \nonumber \\
&&+\frac{\partial d}{\partial \phi }\left. \sin \left( \phi -\phi
_{p}\right) \vphantom{\frac{\partial d}{\partial \eta } }
\right] \,\delta \left( \eta -y\right) d\eta d\phi d\zeta.
\label{dSigma2}
\end{eqnarray}
This form is used in Eq. (\ref{cooperfrye}) to calculate the transverse-momentum spectra. The interesting feature of this formula is that the derivative of the distance $d$ with respect to the spacetime rapidity $\eta$ has disappeared. This means that for each value of the rapidity (and spacetime rapidity) one calculates the spectra using the function $d$ defined exactly for this value of rapidity. Such a function is provided by the hydrodynamic code which should be executed also for exactly the same value of rapidity (treated as a parameter). We thus see that the Cooper-Frye prescription is consistent with the physical picture of non-interacting transversally expanding 2D clusters. Moreover, the system under study is not necessarily boost-invariant.

\section{Results}
\label{sect:results}

In Fig. \ref{fig:pT} we show the PHENIX pion spectra for the centrality classes 20 - 30\% and 60 - 70\% \cite{RHIC-spectra}, while in Fig. \ref{fig:v2} we show the PHENIX data on the elliptic flow for the centrality classes 20 - 40\% and 40 - 60\% \cite{RHIC-v2}. In both cases the data (solid lines) are compared with our hydrodynamic calculations (dashed lines) done for the appropriate values of the impact parameter. Following the PHENIX study performed within the Glauber model \cite{RHIC-spectra} we used the values: \mbox{$b$ = 7.25 fm} for \mbox{$c$ = 20 - 30\%}, \mbox{$b$ = 7.94 fm} for \mbox{$c$ = 20 - 40\%}, \mbox{$b$ = 10.26 fm} for \mbox{$c$ = 40 - 60\%}, and \mbox{$b$ = 11.69 fm} for \mbox{$c$ = 60 - 70\%}. The experimental spectrum of pions shown in Fig. \ref{fig:pT} is the spectrum of positive pions multiplied by a factor of 3. This is done to account for the total hadron multiplicity.

Our results were obtained for the initial central temperature $T_i$ = 250 MeV and for four different values of the final temperature $T_f$ = 200, 180, 160 and 140 MeV.  Similarly to the results obtained by Heinz and Wong we observe that a lower than usual (i.e., lower than for 3D expansion) initial temperature is required to describe correctly the slope of the experimental spectrum. This effect is related to the generation of stronger transverse flow in the case of 2D expansion. In agreement with Ref. \cite{HeinzWong} we also find that the shape of the spectrum is quite insensitive to the final temperature since the lowering of the temperature in the distribution function is compensated by the increase of the magnitude of the transverse flow. This property can be understood in our case as an effect of the exact conservation of the transverse energy and entropy. Since the entropy current is proportional to the particle current, the particle number is also conserved. Combing the conservation laws with the fact that particles have fixed rapidity we find that the average transverse momentum must be exactly conserved
\begin{equation}
\frac{\frac{dE_\perp}{dy}}{\frac{dN}{dy}} = \frac{dE_\perp}{dN} = \langle p_\perp \rangle = \hbox{const}.
\end{equation}
The conservations of the average transverse momentum and the particle number do not allow for substantial changes of the slope of the spectrum. This behavior is shown in Fig. \ref{fig:pT}, where the spectra for different final temperatures are shown to be very much similar.

The results presented in Fig. \ref{fig:pT} were obtained with the normalization factor $n_0$ = 1. The correct normalization for the peripheral collisions may be obtained by the small reduction of $n_0$, while the correct normalization for the central collisions requires larger values of $n_0$. Dependence of the normalization factor on the centrality reflects the experimental fact that the total multiplicity grows faster with centrality than the number of the wounded nucleons. Frequently, this effect is understood as the extra contribution from the binary collisions.  A possible explanation of such a centrality dependence is also offered by the model of wounded quarks and diquarks \cite{bb1,bb2}. Since such effects are not included in the form of our initial condition (\ref{initialeps2}), different values of $n_0$ should be applied for different values of the centrality.

In Fig. \ref{fig:v2} we show our main results concerning the elliptic flow coefficient. The very striking observation is that the elliptic flow is large and for lower values of the final temperature it even exceeds the data. The origin of this effect is the genuine two-dimensional hydrodynamic expansion which transforms the initial thermal energy exclusively into the transverse flow. We observe that for peripheral collisions the elliptic flow coefficient $v_2$ becomes compatible with the data already at the temperature as high as 200 MeV. For more central collisions the experimental values are recovered at smaller but still high values of $T_f$.

\section{Conclusions}

In this paper we have shown that the idea that the system created in ultra-relativistic heavy-ion collisions undergoes the hydrodynamic expansion only in the transverse direction is compatible with the experimental data. With a suitable choice of the initial and final temperatures one is able to describe the measured hadron spectra and the elliptic flow coefficient $v_2$. Clearly, further investigations and developments of such a simple idea should be performed to address more subtle aspects of hadron production. In particular, the present description should be supplemented with the hadronization model. \\   

\noindent {\bf Acknowledgments:} We thank Professor Andrzej Bialas for his inspirations to continue this research and for the critical reading of the manuscript.

\section{Appendix}

The standard definition of the energy-momentum tensor as the second moment of the distribution function in the momentum space yields 
\begin{eqnarray}
T^{\mu \nu } = \frac{n_0 \nu_g T^{\,3}}{2\pi \tau } t^{\mu \nu }, \nonumber
\label{tmunu}
\end{eqnarray}
where (\mbox{$\sinh x$ = sh $x$}, $\cosh x$ = ch $x$)
\begin{eqnarray}
& & \hspace{-0.75cm} t^{\mu \nu} \! = \! \left( 
\begin{array}{cccc}
\hbox{ch} ^{2}\eta \,\left( 2u_{0}^{2}+u_{\bot }^{2}\right) & \hbox{ch} \eta
\,3\,u_{0}u_{x} & \hbox{ch} \eta \,3\,u_{0}u_{y} & \hbox{ch} \eta \hbox{sh} \eta
\,\left( 2u_{0}^{2}+u_{\bot }^{2}\right) \\ 
\hbox{ch} \eta \,3\,u_{0}u_{x} & \,\left( 1+3u_{x}u_{x}\right) & \,3u_{x}u_{y}\,
& \hbox{sh} \eta \,3\,u_{0}u_{x} \\ 
\hbox{ch} \eta \,3\,u_{0}u_{y} & \,3\,u_{y}u_{x} & \,\left( 1+3u_{y}u_{y}\right)
& \hbox{sh} \eta \,3\,u_{0}u_{y} \\ 
\hbox{ch} \eta \hbox{sh} \eta \,\left( 2u_{0}^{2}+u_{\bot }^{2}\right) & \hbox{sh} \eta
\,3\,u_{0}u_{x} & \hbox{sh} \eta \,3\,u_{0}u_{y} & \hbox{sh} ^{2}\eta \,\left(
2u_{0}^{2}+u_{\bot }^{2}\right)
\end{array} \hspace{-0.2cm}
\right) . \nonumber
\label{Tmunu-matrix}
\end{eqnarray}
This expression may be rewritten in the compact form, presented in the lower line in Eq. (\ref{Tmunu}), if the definitions of the four-vectors $U^\mu$  and $V^\mu$ are used.


\end{document}